\documentclass{article}

\usepackage{PRIMEarxiv}

\usepackage[utf8]{inputenc} 
\usepackage[T1]{fontenc}    
\usepackage{hyperref}       
\usepackage{url}            
\usepackage{booktabs}       
\usepackage{amsfonts}       
\usepackage{nicefrac}       
\usepackage{microtype}      
\usepackage{graphicx}       
\graphicspath{{figures/}}   
\usepackage{amsmath}
\usepackage{array}
\usepackage{multirow}
\usepackage{enumitem}
\usepackage{xcolor}
\usepackage{listings}
\usepackage{float}
\usepackage{tikz}
\usetikzlibrary{arrows.meta,backgrounds,shapes.geometric,positioning,fit,calc}
\usepackage{pgfplots}
\pgfplotsset{compat=newest}
\usepgfplotslibrary{fillbetween}
\hypersetup{
  colorlinks=true,
  linkcolor=blue!50!black,
  citecolor=blue!50!black,
  urlcolor=blue!50!black
}

\title{Beyond the Data Mesh Illusion: \\ Designing Modern AI-augmented Lakehouses to \\ Bridge the Gap Between Theory and Practice}

\author{
  Oliver Ang\'elil \\
  ishango.ai \\
  Zurich, Switzerland \\
  \And
  Jan Migon \\
  Independent Researcher \\
  Lausanne, Switzerland \\
}

\begin{document}

\maketitle

\begin{abstract}
Enterprise data platforms face an enduring tension between domain self-service and holistic governance. The data mesh paradigm proposed decentralized domain ownership as a remedy, but pure implementations frequently underdeliver---teams inherit new responsibilities without the platform maturity, tooling, or coordination mechanisms needed to exercise them effectively. This paper argues that the flexibility-versus-control trade-off can be relaxed through an AI-augmented hub-and-spoke model layered on a modern lakehouse architecture. A central hub (Center of Excellence) provides shared platform services, policy automation, and AI-enabled governance---automatically standardizing data products, generating quality rules, drafting data contracts, and reviewing changes for regressions. Domain spokes own business semantics, product backlogs, and local iteration cadence, progressively assuming greater responsibility as they mature. The same LLMs that automate governance tasks also lower the barrier for domain practitioners to develop genuine cross-functional expertise spanning business and data engineering, enabling spoke teams to take on greater end-to-end ownership without proportionally increasing their dependence on the hub. Natural-language conversational interfaces further democratize access for business users, exposing historically underutilized enterprise data. On the organizational side, we propose a staged framework that shifts ownership from hub to spokes, avoiding both centralized bottlenecks and uncoordinated decentralization. We evaluate the architecture through three outcome metrics---data product adoption, time-to-find, and time-to-insight---that tie platform success to measurable business value rather than internal activity.
\end{abstract}


\section{Introduction}
\label{sec:introduction}

Enterprise data platforms are under pressure from two opposing forces. On one side, organizations need rapid delivery of analytics and AI capabilities to support operational decision-making, experimentation, and product development. On the other, they must satisfy non-negotiable requirements around lineage, access control, semantic consistency, data quality, and regulatory compliance. This recurring tension produces a false choice between centralized efficiency and decentralized agility. Centralized data teams often accumulate requests faster than they can deliver them. Fully decentralized operating models can move quickly at first, but they tend to fragment standards, duplicate effort, and erode trust in shared data assets.

Data mesh was proposed as a sociotechnical response to this tension, advancing domain ownership, data as a product, self-serve platform infrastructure, and federated computational governance~\cite{dehghani2022}. The framework is compelling because it reframes data management as an organizational design problem rather than a purely technical one. Yet many real-world implementations have struggled to operationalize the model at scale. The language of decentralization is frequently adopted without the sustained investment required in platform abstractions, governance processes, and product management capabilities~\cite{thoughtworks2026}. A central IT function that continues to build and maintain domain pipelines on behalf of business units---because those units lack the engineering capacity to do so---has not implemented data mesh; it has merely distributed a traditional centralized architecture without transferring the underlying responsibility. In consequence, teams inherit additional responsibilities without gaining the engineering capacity, tooling, incentives, or cross-domain coordination needed to perform them effectively.

\begin{figure}[H]
  \centering
  \definecolor{mcknav}{RGB}{0,42,87}       
  \definecolor{mckblue}{RGB}{0,112,185}    
  \definecolor{mckred}{RGB}{196,18,48}     
  \definecolor{mckgreen}{RGB}{0,121,64}    
  \definecolor{mckgray}{RGB}{89,89,89}     
  \definecolor{mckfill}{RGB}{245,238,232}  
  \begin{tikzpicture}[x=1cm, y=1cm, scale=1.35]
    \fill[mckfill]
      (0, {12/1.3}) --
      plot[domain=1.3:9.2, smooth, samples=80] (\x, {12/\x}) --
      (9.2, 0) -- (0, 0) -- cycle;
    \draw[mckred, line width=2.8pt, dash pattern=on 5pt off 3pt]
      plot[domain=1.3:9.2, smooth, samples=80] (\x, {12/\x});
    \draw[-{Latex[length=4mm, width=3mm]}, mckblue, line width=2pt, dashed]
      (0.2, 0.2) -- (4.55, 4.55);
    \filldraw[mcknav] (1.8, 7.2) circle (3pt);
    \node[font=\large\bfseries, text=mcknav] at (2.2, 7.25) {A};
    \node[font=\footnotesize, text=mckgray, align=left, anchor=west]
      at (2.6, 7.2) {Bottlenecked\\central team};
    \filldraw[mcknav] (8.0, 1.4) circle (3pt);
    \node[font=\large\bfseries, text=mcknav] at (8.05, 1.75) {B};
    \node[font=\footnotesize, text=mckgray, align=left, anchor=west]
      at (7.4, 2.35) {Fragmented\\domains};
    \node[star, star points=5, star point ratio=2.2,
          fill=mckgreen, draw=mckgreen!80!black,
          minimum size=9mm, line width=0.5pt] at (5.0, 5.0) {};
    \node[font=\small\bfseries, text=mckgreen, align=left, anchor=west]
      at (5.65, 5.05) {Proposed\\model};
    \node[font=\small\bfseries, text=mckblue, align=center]
      at (7, 6.5)
      {AI-augmented hub-and-spoke\\can break the traditional\\trade-off frontier};
    \node[font=\small\bfseries\itshape, text=mckred]
      at (5.0, 0.5) {traditional trade-off frontier};
    \draw[mcknav, line width=2pt, -{Latex[length=4mm, width=3mm]}]
      (0, 0) -- (10.7, 0);
    \draw[mcknav, line width=2pt, -{Latex[length=4mm, width=3mm]}]
      (0, 0) -- (0, 10.7);
    \node[font=\small\bfseries, text=mcknav] at (5.35, -0.45)
      {Data Governance (\textquotedblleft control\textquotedblright)};
    \node[font=\small\bfseries, text=mcknav, rotate=90] at (-0.45, 5.35)
      {Domain Autonomy (\textquotedblleft flexibility\textquotedblright)};
  \end{tikzpicture}
  \caption{Traditional trade-off between governance control and domain flexibility. Point~A represents the status quo in many organizations: a central data-team bottleneck pushes frustrated domain teams toward local autonomy, but standards are applied unevenly and teams drift into silos. Point~B represents the opposite failure mode, where a central team enforces guardrails so rigidly that domain teams lose practical flexibility and delivery slows. A well-designed sociotechnical model aims to move beyond this frontier by increasing both control and flexibility simultaneously.}
  \label{fig:frontier}
\end{figure}
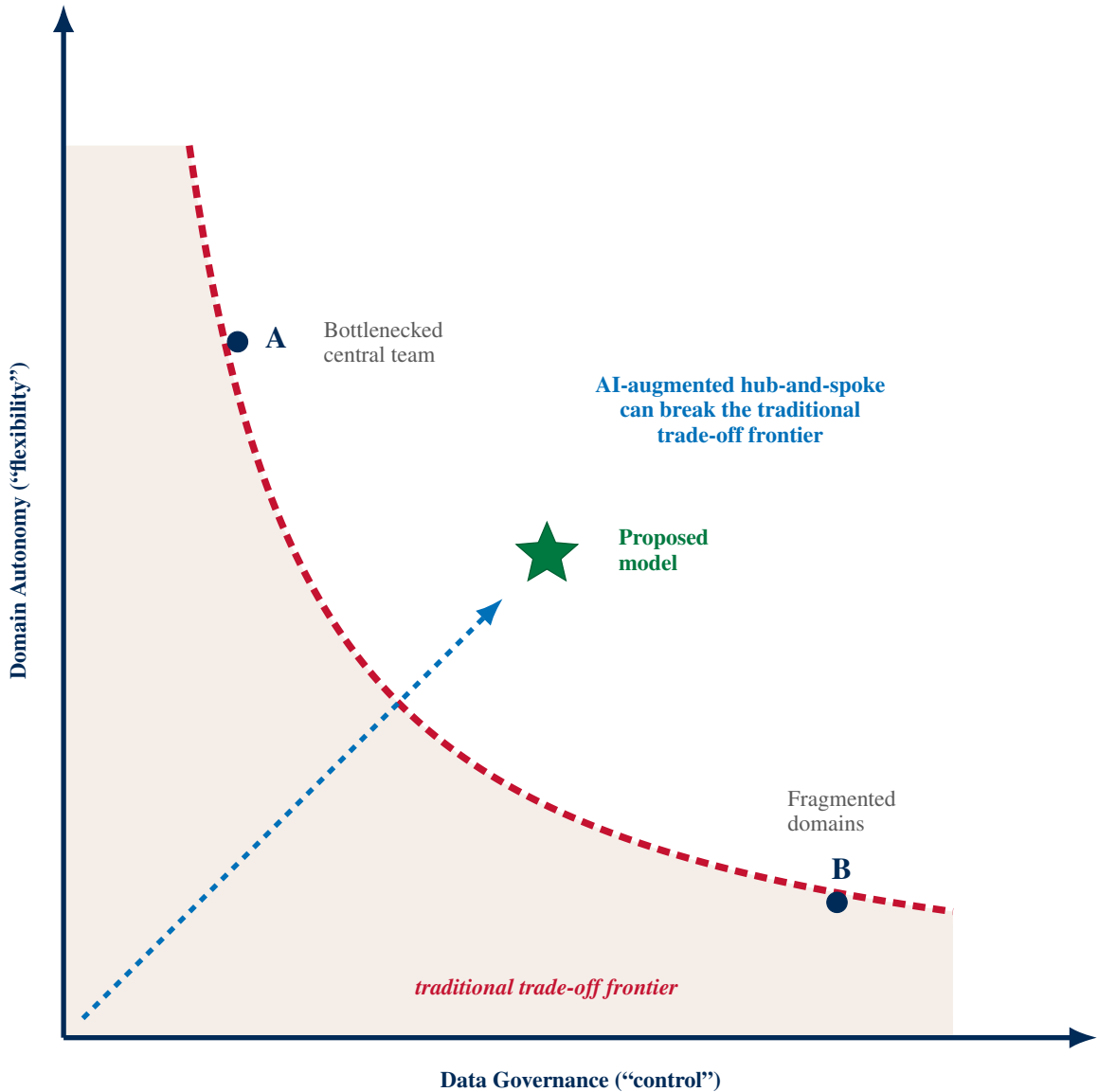

Figure~\ref{fig:frontier} visualizes the traditional frontier that many enterprises experience in practice. As governance control becomes more centralized and prescriptive, domain teams often lose the flexibility to publish, refine, and consume data products in the cadence their business context demands. When that flexibility is expanded without a strong shared control plane, the result is usually fragmented standards rather than productive autonomy. The design problem is therefore not merely to choose a point on the frontier, but to reshape it through a sociotechnical model that lowers the coordination cost of governance.

This paper argues that most practical failures stem not from rejecting data mesh principles but from assuming that pure decentralization is sufficient to realize them. We propose that the flexibility-versus-control trade-off can be relaxed more effectively through a hub-and-spoke operating model layered on top of an AI-augmented modern lakehouse architecture. In this model, a central hub provides platform services, policy automation, interoperability standards, and governance tooling, while business-aligned spokes own domain semantics, operational priorities, and local product backlogs. Artificial intelligence is used not as a substitute for governance, but as a force multiplier for it: automating repetitive standardization work, generating metadata and data contracts, reviewing product changes, and exposing enterprise knowledge through conversational discovery interfaces.

Our contribution is threefold. First, we synthesize the practical limitations of pure data mesh through the lens of organizational coordination and platform maturity. Second, we define a hub-and-spoke architecture in which AI augments both domain autonomy and central stewardship. Third, we propose an evaluation framework grounded in measurable outcomes, including data product usage, time-to-find, and time-to-insight. The resulting model aims to preserve the spirit of domain-oriented data ownership while remaining compatible with the operational realities of modern enterprises.

\section{Related Work and Background}
\label{sec:related}

This section surveys the key intellectual threads that motivate our proposal. We organize the discussion thematically, beginning with data mesh itself, then moving through lakehouse architectures, governance challenges, organizational design, AI-enabled coordination, and adjacent paradigms.

\begin{enumerate}[leftmargin=1.5em, label=\textbf{\arabic*.}]

  \item \textbf{Data mesh foundations.} Dehghani formalized the four core principles of domain-oriented ownership, data as a product, self-serve platform, and federated governance~\cite{dehghani2019,dehghani2022}. The core insight is that data bottlenecks mirror the limitations of monolithic software organizations: scaling analytics and machine learning requires distributing responsibility closer to the teams that generate and understand the data.

  \item \textbf{Lakehouse architectures.} The modern lakehouse attempts to combine the flexibility and cost efficiency of data lakes with the ACID guarantees, schema enforcement, and management capabilities of warehouses~\cite{armbrust2021}. Lakehouse architectures offer standardized table formats, centralized cataloging, and strong lineage capabilities that make them a promising substrate for product-oriented data sharing. They also provide a practical convergence point for batch analytics, streaming ingestion, and machine learning workflows.

  \item \textbf{Persistent governance challenges.} A lakehouse alone does not solve the governance problem. Because lakehouses evolved from---and still share core infrastructure with---data lakes, they inherit many of the same challenges: metadata sparsity, schema evolution conflicts, and discoverability difficulties that persist even when the underlying storage is well managed~\cite{nargesian2020}. Schema mismatches in particular surface only at runtime---columns dropped, types changed silently between pipeline nodes---and remain a leading cause of production incidents even on well-governed lakehouses~\cite{bauplan2026}. FAIR data principles emphasize that data must be findable, accessible, interoperable, and reusable to support meaningful downstream use~\cite{wilkinson2016}. Those properties depend on persistent metadata, common semantics, access management, and discoverability workflows. Prior work on data contracts further suggests that trustworthy data sharing requires explicit schema and behavioral guarantees between producers and consumers~\cite{breck2019}. Without those guarantees, domain autonomy often devolves into incompatible local optimizations.

  \item \textbf{Socio-technical systems and platform teams.} The literature on team topologies and platform engineering cautions against simplistic decentralization. Team cognitive load, interface quality, and organizational topology directly shape delivery outcomes~\cite{skelton2019}. Where decentralization succeeds, it is usually supported by a strong enabling platform function and clear ownership boundaries rather than by the disappearance of centralized capability. Industry experience confirms this: ``decentralised'' data mesh pilots often reveal persistent dependency on a central team for common models, infrastructure maintenance, security review, and metadata stewardship~\cite{thoughtworks2026}. Practitioner-led implementations on modern lakehouse platforms formalize this into two practical enterprise topologies that sit at different points on a centralization spectrum. A \textit{harmonized data mesh} is a pragmatic middle ground: domains still share data products peer-to-peer, but a common catalog and shared infrastructure blueprints provide security, compliance, and discoverability across them---distinguishing it from fully decentralized models where each domain manages its own infrastructure in isolation. A \textit{hub-and-spoke data mesh} goes further, routing data products through a central hub that actively manages shared assets and provides platform services to each domain spoke. The harmonized approach is noted to be challenging in global organizations with uneven engineering depth, where teams struggle to stay independently in sync with shared standards~\cite{walter2022}. Regardless of which topology is chosen, a recurring pattern is that organizations adopt the vocabulary of data mesh without investing in the platform abstractions, incentive structures, and product management discipline that either model presupposes.

  \item \textbf{AI-enabled coordination.} Recent advances in large language models change the economics of governance work. These models can assist with metadata authoring, schema mapping, policy review, and natural-language access to data assets~\cite{narayan2022}. Used carefully, they reduce the manual overhead of documentation and governance while keeping humans accountable for consequential decisions. Early work on natural-language interfaces to databases suggests that non-expert users can query structured data productively when the interface is grounded in well-maintained metadata~\cite{li2024}. This opens a middle path: AI-assisted platform services can take on much of the coordination work that has historically required either a large central team or strong domain discipline - potentially making decentralization more viable without abandoning governance quality.

  \item \textbf{Conway's Law and organizational architecture.} System architectures tend to mirror the communication structures of the organizations that build them~\cite{conway1968}. If teams are siloed, data will be too, not because of bad technical decisions, but because people naturally build to their own boundaries. This means that fixing a fragmented data architecture requires changing how teams are organized, not just how data is stored or processed. The CoE model we propose is an attempt to do that deliberately: to define team boundaries and ownership in a way that makes coherent, shareable data the path of least resistance.

  \item \textbf{Dark data.} The concept of ``dark data''---enterprise information that is collected and stored but never analyzed or leveraged---has been documented as a significant source of untapped value~\cite{heidorn2008}. Estimates suggest that a majority of enterprise data falls into this category. Making dark data discoverable and usable is a practical priority that motivates the conversational access layer in our architecture.

  \item \textbf{Data contracts.} The data contracts discourse has matured considerably beyond early data validation frameworks~\cite{breck2019}. Recent practitioner-led specifications define contracts as formal agreements between data producers and consumers, encompassing schema guarantees, freshness expectations, and semantic definitions~\cite{jones2023}. In the systems literature, the schema dimension is addressed through typed table contracts that make pipeline interfaces machine-checkable at the boundary between producer and consumer~\cite{bauplan2026}. These contracts serve as the enforceable interface between autonomous domains and are central to our proposed governance automation.

\end{enumerate}

\section{Methods: An AI-Augmented Hub-and-Spoke Model}
\label{sec:methods}

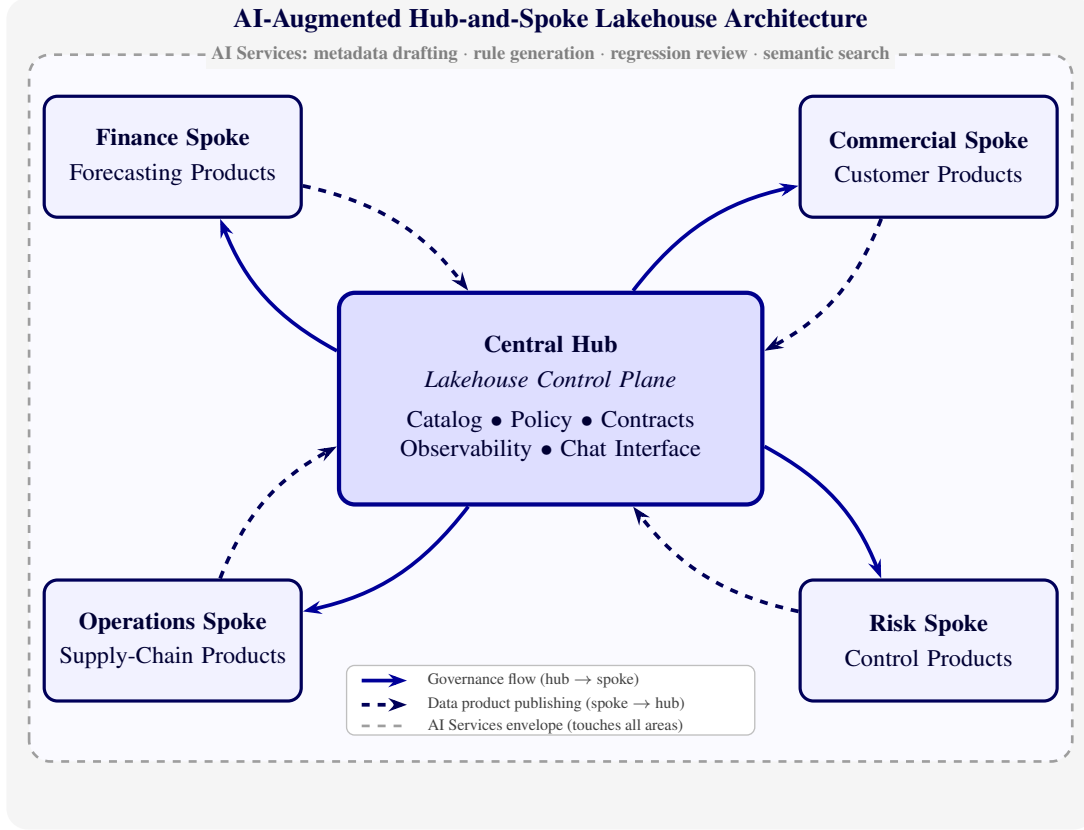
\begin{figure}[!ht]
  \centering
  \begin{tikzpicture}[
    >=Stealth,
    every node/.style={font=\small},
    hub/.style={
      rectangle, rounded corners=6pt,
      draw=blue!55!black, line width=1.6pt, fill=blue!13!white,
      minimum width=5.6cm, minimum height=2.8cm,
      text width=5.2cm, align=center, text=blue!20!black
    },
    spoke/.style={
      rectangle, rounded corners=5pt,
      draw=blue!35!black, line width=1.2pt, fill=blue!5!white,
      minimum width=3.4cm, minimum height=1.6cm,
      text width=3.0cm, align=center, text=blue!20!black
    },
    govflow/.style={
      -{Stealth[length=7pt,width=5pt]},
      line width=1.4pt, blue!60!black
    },
    dataflow/.style={
      -{Stealth[length=7pt,width=5pt]},
      line width=1.4pt, blue!35!black, dashed
    },
  ]
    \begin{scope}[on background layer]
      \fill[black!4!white, rounded corners=8pt]
          (-7.2, -5.7) rectangle (7.2, 5.3);
    \end{scope}

    \begin{scope}[on background layer]
      \filldraw[rounded corners=8pt,
                draw=black!38, line width=1.0pt, dashed,
                fill=blue!2!white]
          (-6.9, -4.8) rectangle (6.9, 4.55);
    \end{scope}

    \node[font=\normalsize\bfseries, text=blue!25!black] at (0, 5.0)
      {AI-Augmented Hub-and-Spoke Lakehouse Architecture};

    \node[font=\scriptsize\bfseries, text=black!50,
          fill=black!4!white, inner sep=2pt]
      at (0, 4.55) {\textbf{AI Services:} metadata drafting $\cdot$
                    rule generation $\cdot$ regression review $\cdot$
                    semantic search};

    \node[hub] (hub) at (0, 0) {
      \textbf{Central Hub}\\[3pt]
      \textit{Lakehouse Control Plane}\\[5pt]
      \footnotesize Catalog $\bullet$ Policy $\bullet$ Contracts\\[1pt]
      \footnotesize Observability $\bullet$ Chat Interface
    };

    \node[spoke] (finance)    at (-5.0,  3.2)
      {\textbf{Finance Spoke}\\[3pt]\footnotesize Forecasting Products};
    \node[spoke] (commercial) at ( 5.0,  3.2)
      {\textbf{Commercial Spoke}\\[3pt]\footnotesize Customer Products};
    \node[spoke] (operations) at (-5.0, -3.2)
      {\textbf{Operations Spoke}\\[3pt]\footnotesize Supply-Chain Products};
    \node[spoke] (risk)       at ( 5.0, -3.2)
      {\textbf{Risk Spoke}\\[3pt]\footnotesize Control Products};

    \draw[govflow,  bend left=20] (hub) to (finance);
    \draw[govflow,  bend left=20] (hub) to (commercial);
    \draw[govflow,  bend left=20] (hub) to (operations);
    \draw[govflow,  bend left=20] (hub) to (risk);

    \draw[dataflow, bend left=20] (finance)    to (hub);
    \draw[dataflow, bend left=20] (commercial) to (hub);
    \draw[dataflow, bend left=20] (operations) to (hub);
    \draw[dataflow, bend left=20] (risk)       to (hub);

    \begin{scope}[shift={(0, -4.15)}]
      \begin{scope}[on background layer]
        \filldraw[fill=white, draw=black!25, rounded corners=3pt, line width=0.5pt]
          (-2.7, -0.30) rectangle (2.7, 0.62);
      \end{scope}
      \draw[govflow]  (-2.50, 0.42) -- (-1.90, 0.42);
      \node[anchor=west, font=\tiny, text=black!75] at (-1.75, 0.42)
        {Governance flow (hub $\to$ spoke)};
      \draw[dataflow] (-2.50, 0.10) -- (-1.90, 0.10);
      \node[anchor=west, font=\tiny, text=black!75] at (-1.75, 0.10)
        {Data product publishing (spoke $\to$ hub)};
      \draw[black!38, dashed, line width=1.0pt] (-2.50, -0.18) -- (-1.90, -0.18);
      \node[anchor=west, font=\tiny, text=black!75] at (-1.75, -0.18)
        {AI Services envelope (touches all areas)};
    \end{scope}
  \end{tikzpicture}
  \caption{AI-augmented hub-and-spoke lakehouse architecture. The central hub provides the
    shared control plane, while domain spokes own business-facing data products. Solid
    arrows indicate governance flows (hub $\to$ spokes); dashed arrows show data product
    publishing (spokes $\to$ hub). The outer dashed border marks the AI Services layer,
    which supports all areas through metadata drafting, data contract generation
    and chat interfaces.}
  \label{fig:architecture}
\end{figure}

Our proposal pairs a lakehouse-centered technical architecture with an operating model that combines central platform stewardship and decentralized domain ownership (Figure~\ref{fig:architecture}). The central hub---the Center of Excellence (CoE)---owns the shared control plane: metadata catalog, policy engine, data contract generator, contract registry, chat interface, and observability infrastructure. Domain teams (spokes) own business semantics and author their own metadata entries within that catalog.

This topology goes beyond the harmonized approach, in which domains share infrastructure peer-to-peer without a central accountable hub, and beyond the uncoordinated independence of a pure decentralized mesh~\cite{walter2022}. The CoE remains explicitly accountable for standards, enablement, and platform reliability---reducing coordination costs without recentralizing data product ownership.

Technically, the model is built around five capabilities: four AI-enabled methods and one shared lakehouse substrate.

\begin{enumerate}[leftmargin=1.5em]
  \item \textbf{AI-assisted data product documentation.} Each domain team publishes data products with metadata and documentation that meet CoE standards. AI assistance is used to draft initial metadata, propose documentation, and infer upstream dependencies from transformation code. For example, when a commercial domain publishes a campaign-performance mart, an assistant can draft the product summary and identify upstream sources from the underlying SQL. The producer approves the final product definition, but the administrative burden of publication falls sharply.
  \item \textbf{AI-generated data contracts} Data contracts translate informal expectations about a data product into explicit, executable guarantees covering schema, quality rules, and behavioral commitments between producers and consumers. Rather than requiring producers to author these from scratch, AI can draft a contract by reasoning over the dataset's schema and metadata, applicable compliance requirements, and any business context the producer supplies. This includes proposing quality rules on field constraints, acceptable value ranges, and timeliness that reflect intent and compliance context rather than just statistical observation. The producer and relevant governance stakeholders review and approve the result before it is registered. The value is not that AI decides policy but that it produces a credible first version, grounded in real context, for human ratification.
  \item \textbf{AI-assisted data profiling for security.} Data products are monitored by automated review agents that autonomously scan incoming data values to flag any sensitive information. For example, AI agents can detect personally identifiable information (PII) such as names, addresses, or credit card numbers in unstructured or semi-structured fields. When sensitive data is identified, the system can automatically apply appropriate privacy classifications, trigger encryption or masking protocols, and route the findings to security approvers---embedding governance directly into the delivery workflow rather than deferring it until after a potential incident.
  \item \textbf{Conversational discovery and access.} Enterprise users interact with cataloged assets through an agentic interface that translates natural-language questions into governed retrieval and analysis. The assistant reasons over metadata that domain owners have authored and published, including definitions, lineage and business context, rather than over raw data directly. Access control is enforced at the platform level and the assistant cannot circumvent it: a user sees only what they are permitted to see, and queries are executed within those boundaries regardless of how the question is phrased. What distinguishes this from standard search is that the assistant does not return a list of datasets but attempts to answer the underlying business question, surfacing relevant products, explaining key definitions, and offering interpretive context about why a metric looks the way it does. A supply-chain manager asking why late shipments rose in warehouse 17 last month receives not a pointer to a logistics table but an explanation grounded in certified data, with the reasoning made visible.
  \item \textbf{Shared lakehouse substrate.} Shared storage formats, centralized metadata, and lineage services provide the common substrate required for reuse across domains~\cite{armbrust2021,walter2022}. This allows spokes to move independently while remaining legible to the rest of the enterprise. AI-enabled workflows become more reliable when they operate over this common substrate because lineage, schema history, and access policy are machine-readable rather than tribal knowledge.
\end{enumerate}

Among these methods, AI-powered data contracts are especially important because they turn informal expectations into executable publication checks. Recent structured-output capabilities in large language models are useful here: instead of generating free-form governance prose, the model can be constrained to emit a typed contract object that conforms to a predefined schema before it ever reaches a registry or CI pipeline. That substantially reduces parsing ambiguity and makes review workflows more practical.
Figure~\ref{fig:c4} illustrates the full pipeline. Three inputs — a metadata fetcher, a compliance loader, and a free-text intake — collect schema definitions, regulatory rules, and business annotations, forming the context layer that grounds the model in dataset-specific facts. The LLM orchestrator assembles these into a structured prompt and calls a foundation model constrained to emit a typed contract object. The output is verified by a data product owner before being persisted as a versioned YAML or JSON file in the contract store.\footnote{A working sample Python implementation of this pattern is available at \url{https://github.com/JanMigon/ai-data-contract-generator/blob/main/example.py}.}

\definecolor{clrGray}{RGB}{220,218,210}
\definecolor{clrPurple}{RGB}{206,203,246}
\definecolor{clrPurpleDark}{RGB}{83,74,183}
\definecolor{clrTeal}{RGB}{159,225,203}
\definecolor{clrTealDark}{RGB}{15,110,86}
\definecolor{clrAmber}{RGB}{250,199,117}
\definecolor{clrAmberDark}{RGB}{133,79,11}
\definecolor{clrCoral}{RGB}{245,196,179}
\definecolor{clrCoralDark}{RGB}{153,60,29}
\definecolor{clrBorder}{RGB}{160,158,150}
\definecolor{clrArrow}{RGB}{100,98,90}

\tikzset{
  c4node/.style={
    rectangle, rounded corners=4pt,
    draw=clrBorder, line width=0.5pt,
    text width=#1, align=center,
    inner sep=6pt, minimum height=44pt,
    font=\small
  },
  c4node/.default=90pt,
  gray node/.style={c4node=#1, fill=clrGray},
  gray node/.default=90pt,
  purple node/.style={c4node=#1, fill=clrPurple},
  purple node/.default=90pt,
  teal node/.style={c4node=#1, fill=clrTeal},
  teal node/.default=90pt,
  amber node/.style={c4node=#1, fill=clrAmber},
  amber node/.default=90pt,
  coral node/.style={c4node=#1, fill=clrCoral},
  coral node/.default=90pt,
  ntitle/.style={font=\small\bfseries},
  nsub/.style={font=\scriptsize\color{black!70}},
  c4arrow/.style={
    -{Stealth[length=5pt,width=4pt]},
    draw=clrArrow, line width=0.8pt
  },
  seclabel/.style={font=\scriptsize\color{black!45}, anchor=west}
}

\providecommand{\nbox}[2]{\textbf{#1}\\[2pt]\scriptsize\color{black!65}#2}

\begin{figure}[H]
\centering

\begin{tikzpicture}[
  node distance = 0pt,   
  remember picture
]


\draw[clrBorder, line width=0.4pt] (0, 1.1) -- (16.0, 1.1);
\node[seclabel] at (0, 1.28) {C1 · System context};

\node[gray node=65pt]    (eng) at (1.30, -0.1)
      {\nbox{Data engineer}{Triggers contract creation}};

\node[purple node=100pt] (sys) at (7.20, -0.1)
      {\nbox{Contract creation system}{LLM assisted}};

\node[gray node=65pt]    (con) at (13.50, -0.1)
      {\nbox{Data consumer}{Reads \& validates}};

\draw[c4arrow] (eng.east) -- node[above, font=\scriptsize, text=clrArrow]{requests} (sys.west);
\draw[c4arrow] (sys.east) -- node[above, font=\scriptsize, text=clrArrow]{exports}  (con.west);



\draw[clrBorder, line width=0.4pt] (0, -1.78) -- (16.0, -1.78);
\node[seclabel] at (0, -1.60) {C2 · Containers};

\draw[clrTealDark, line width=0.7pt, dashed, rounded corners=5pt]
     (-0.75, -2.13) rectangle (3.85, -10.05);
\node[font=\scriptsize\bfseries, text=clrTealDark, anchor=north west]
     at (-0.60, -2.13) {Context Engineering};

\node[teal node=90pt, minimum height=56pt] (meta) at (1.50, -3.55)
      {\nbox{Metadata fetcher}{Pulls schema, lineage,\\data catalogue entries}};

\node[teal node=90pt, minimum height=56pt] (comp) at (1.50, -5.95)
      {\nbox{Compliance loader}{Fetches GDPR, PII rules,\\retention policies}};

\node[teal node=90pt, minimum height=56pt] (free) at (1.50, -8.35)
      {\nbox{Free-text intake}{Business rules, SLAs,\\owner annotations}};

\node[purple node=110pt, minimum height=56pt] (orch) at (8.00, -5.95)
      {\nbox{LLM orchestrator}{Prompt builder + Claude /\\OpenAI API call}};

\coordinate (valwest)  at (13.80, -5.95);
\coordinate (valsouth) at (14.20, -6.60);

\node[font=\small\bfseries, text=black, anchor=south] at (14.20, -5.20) {Data Product Owner};

\draw[clrAmberDark, line width=0.8pt, fill=clrAmber]
     (14.20, -5.50) circle (0.17cm);

\draw[clrAmberDark, line width=0.8pt]
     (14.20, -5.67) -- (14.20, -6.10);

\draw[clrAmberDark, line width=0.8pt]
     (13.95, -5.82) -- (14.45, -5.82);

\draw[clrAmberDark, line width=0.8pt]
     (14.20, -6.10) -- (13.98, -6.55);

\draw[clrAmberDark, line width=0.8pt]
     (14.20, -6.10) -- (14.42, -6.55);

\node[coral node=90pt, minimum height=56pt] (store) at (14.20, -8.70)
      {\nbox{Contract store}{YAML / JSON files,\\git-versioned}};

\draw[c4arrow] (meta.east)  -| ([xshift=-2pt]orch.west);
\draw[c4arrow] (comp.east)  -- ([xshift=-2pt]orch.west);
\draw[c4arrow] (free.east)  -| ([xshift=-2pt]orch.west);
\node[font=\scriptsize, text=clrArrow, anchor=west] at (3.95, -3.25) {metadata};
\node[font=\scriptsize, text=clrArrow, anchor=west] at (3.95, -5.75) {compliance};
\node[font=\scriptsize, text=clrArrow, anchor=west] at (3.95, -8.15) {requirements};

\draw[c4arrow] (orch.east) -- node[above, font=\scriptsize, text=clrArrow]{structured JSON} (valwest);

\draw[c4arrow] (valsouth) -- node[right, font=\scriptsize, text=clrArrow]{validated data contract} (store.north);


\draw[clrBorder, line width=0.4pt] (0, -10.51) -- (16.0, -10.51);
\node[seclabel] at (0, -10.33) {C3 · LLM orchestrator components};

\node[purple node=90pt] (ctx)  at (2.00, -11.39)
      {\nbox{Context builder}{Assembles system prompt}};

\node[purple node=90pt] (mcal) at (8.00, -11.39)
      {\nbox{Model caller}{Retries, timeout logic}};

\node[purple node=90pt] (sch)  at (14.00, -11.39)
      {\nbox{Schema enforcer}{Tool-use / JSON mode}};

\draw[c4arrow] (ctx.east)  -- (mcal.west);
\draw[c4arrow] (mcal.east) -- (sch.west);


\draw[clrBorder, line width=0.4pt] (0, -12.62) -- (16.0, -12.62);
\node[seclabel] at (0, -12.44) {Output · Data contract fields};

\foreach \cx/\title/\sub in {
  1.13/dataset\_name/string,
  4.33/schema/list[Field],
  7.53/sla/{freshness, quality},
  10.73/compliance/{pii\_fields, gdpr},
  13.93/owner/{team, email}
}{
  \node[coral node=64pt, minimum height=38pt] at (\cx, -13.42)
        {\nbox{\texttt{\title}}{\sub}};
}


\draw[clrBorder, line width=0.4pt] (0, -14.55) -- (16.0, -14.55);

\foreach \x/\col/\label in {
  0.0/clrTeal/Input sources,
  3.2/clrPurple/LLM orchestration,
  7.2/clrAmber/Validation,
  9.8/clrCoral/Output \& storage
}{
  \fill[\col, draw=clrBorder, line width=0.4pt, rounded corners=2pt]
       (\x, -15.01) rectangle (\x+0.30, -14.71);
  \node[font=\scriptsize, anchor=west] at (\x+0.40, -14.86) {\label};
}

\end{tikzpicture}

\caption{C4 architecture diagram of the AI-powered data contract creation system.
Boxes are coloured by role: \textcolor{clrTealDark}{\textbf{teal}} = input sources,
\textcolor{clrPurpleDark}{\textbf{purple}} = LLM orchestration,
\textcolor{clrAmberDark}{\textbf{amber}} = human approval (Data Product Owner),
\textcolor{clrCoralDark}{\textbf{coral}} = output and storage.
A working sample Python implementation of this pipeline is available at
\url{https://github.com/JanMigon/ai-data-contract-generator/blob/main/example.py}.}
\label{fig:c4}
\end{figure}
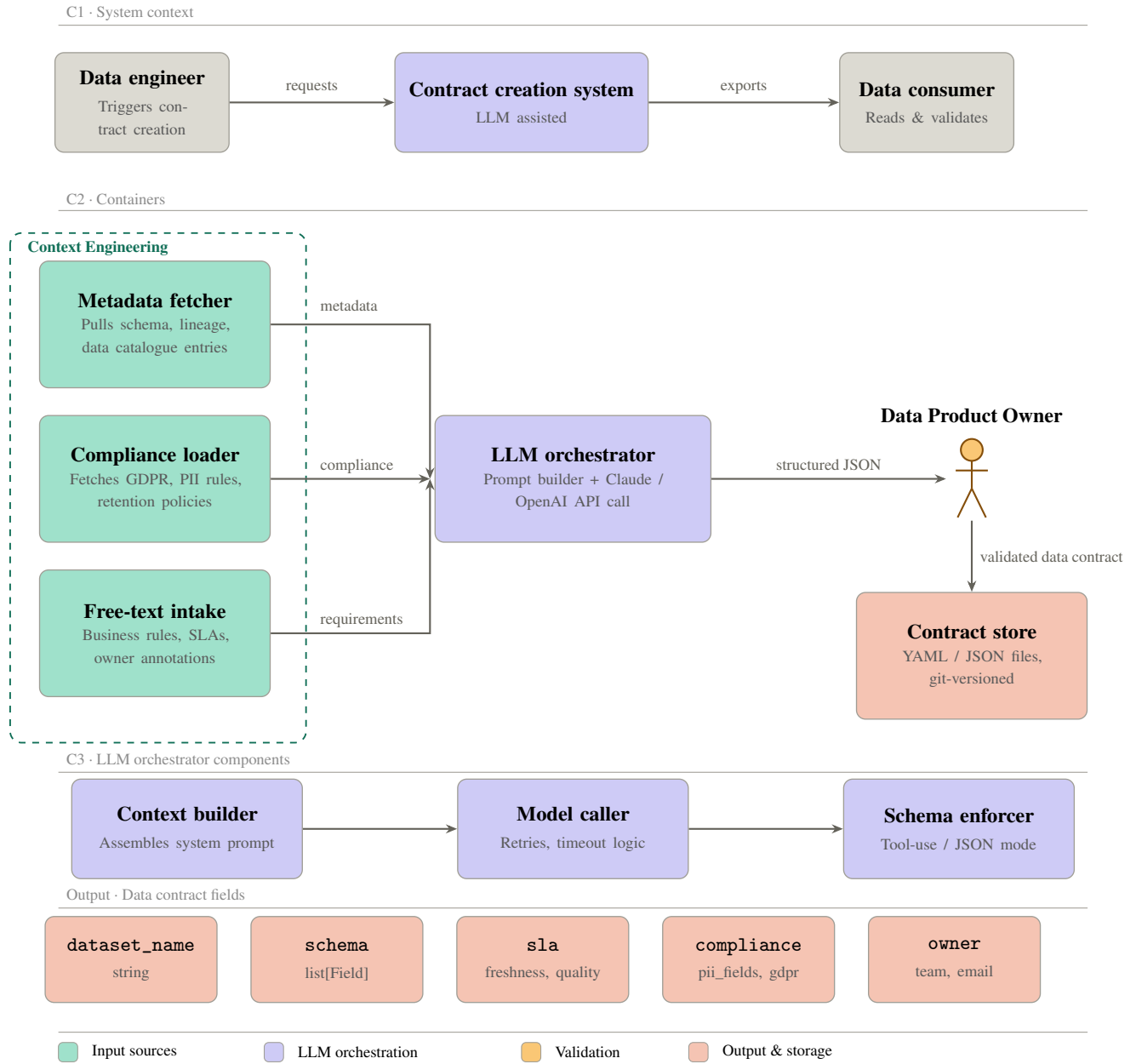

The social operating model is equally important. The CoE is not a command-and-control bottleneck; it provides templates, automation, education, and architectural stewardship. Responsibility migrates outward over time as domain teams demonstrate maturity. Figure~\ref{fig:responsibility} illustrates this progression from hub-led onboarding to domain-led stewardship. We formalize this progression as a staged framework:

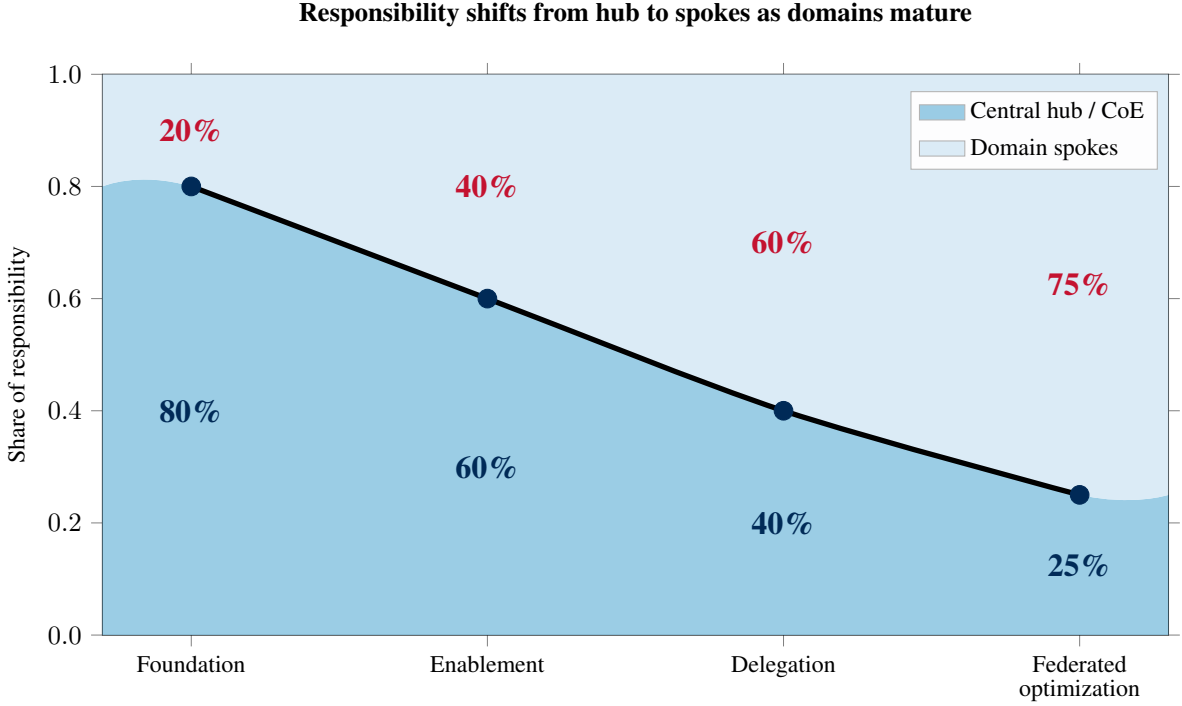
\begin{figure}[H]
  \centering
  \definecolor{hubblue}{RGB}{137,196,225}    
  \definecolor{spokewheat}{RGB}{213,232,245} 
  \definecolor{hubtext}{RGB}{0,42,87}        
  \definecolor{spoketext}{RGB}{196,18,48}    
  \definecolor{hubdot}{RGB}{0,42,87}         
  \begin{tikzpicture}
  \begin{axis}[
    width=0.95\linewidth,
    height=9cm,
    xmin=-0.3, xmax=3.3,
    ymin=0, ymax=1.0,
    xtick={0,1,2,3},
    xticklabels={Foundation, Enablement, Delegation,
                 {\shortstack{Federated\\optimization}}},
    x tick label style={align=center, font=\small},
    ytick={0, 0.2, 0.4, 0.6, 0.8, 1.0},
    y tick label style={/pgf/number format/.cd,
                        fixed, fixed zerofill, precision=1},
    ylabel={Share of responsibility},
    ylabel style={font=\small},
    title={\textbf{Responsibility shifts from hub to spokes as domains mature}},
    title style={font=\normalsize, at={(0.5,1.04)}, anchor=south},
    legend style={
      at={(0.99,0.97)}, anchor=north east,
      font=\small, draw=gray!60,
      fill=white, fill opacity=0.95, text opacity=1,
      row sep=2pt,
    },
    legend cell align=left,
    tick align=outside,
    clip=true,
  ]
    \addplot[name path=hubcurve, smooth, draw=none, forget plot]
      coordinates {(-0.3,0.80) (0,0.80) (1,0.60) (2,0.40) (3,0.25) (3.3,0.25)};
    \addplot[name path=bottom, draw=none, forget plot]
      coordinates {(-0.3,0) (3.3,0)};
    \addplot[name path=topline, draw=none, forget plot]
      coordinates {(-0.3,1) (3.3,1)};
    \addplot[fill=hubblue, fill opacity=0.85, draw=none, forget plot]
      fill between[of=hubcurve and bottom];
    \addplot[fill=spokewheat, fill opacity=0.85, draw=none, forget plot]
      fill between[of=topline and hubcurve];
    \addplot[smooth, black, line width=2pt, forget plot]
      coordinates {(0,0.80) (1,0.60) (2,0.40) (3,0.25)};
    \addplot[only marks, mark=*, forget plot,
             mark options={fill=hubdot, draw=hubdot}, mark size=3.5pt]
      coordinates {(0,0.80) (1,0.60) (2,0.40) (3,0.25)};
    \addlegendimage{fill=hubblue, fill opacity=0.85, draw=none, area legend}
    \addlegendentry{Central hub / CoE}
    \addlegendimage{fill=spokewheat, fill opacity=0.85, draw=none, area legend}
    \addlegendentry{Domain spokes}
    \node[font=\large\bfseries, text=hubtext] at (axis cs:0,   0.400) {80\%};
    \node[font=\large\bfseries, text=hubtext] at (axis cs:1,   0.300) {60\%};
    \node[font=\large\bfseries, text=hubtext] at (axis cs:2,   0.200) {40\%};
    \node[font=\large\bfseries, text=hubtext] at (axis cs:3,   0.125) {25\%};
    \node[font=\large\bfseries, text=spoketext] at (axis cs:0,   0.900) {20\%};
    \node[font=\large\bfseries, text=spoketext] at (axis cs:1,   0.800) {40\%};
    \node[font=\large\bfseries, text=spoketext] at (axis cs:2,   0.700) {60\%};
    \node[font=\large\bfseries, text=spoketext] at (axis cs:3,   0.625) {75\%};
  \end{axis}
  \end{tikzpicture}
  \caption{Responsibility shifts from hub to spokes as domains mature. The CoE retains platform integrity and governance automation, while spokes progressively assume greater ownership of their data products.}
  \label{fig:responsibility}
\end{figure}

\begin{enumerate}[leftmargin=1.5em]
  \item \textbf{Foundation stage.} The CoE defines the minimum viable platform, metadata standards, contract templates, and policy model. Domain teams participate as design partners.
  \item \textbf{Enablement stage.} Spokes begin shipping data products using platform guardrails. AI services reduce the effort required for documentation, schema standardization, and testing. Responsibility for data ingestion rests with the spokes: each domain authors, schedules, and maintains its own pipelines from source systems rather than delegating that work to the CoE. During this stage, the hub accelerates spoke readiness by providing ingestion blueprints and architecture guidance, and by conducting pull request reviews on pipeline code to keep patterns consistent and reusable across domains.
  \item \textbf{Delegation stage.} Mature domains gain expanded control over quality thresholds, release cadence, and local vocabularies, while the hub monitors cross-domain consistency through automated review.
  \item \textbf{Federated optimization stage.} Governance becomes increasingly computational: the hub publishes policies and observability expectations, while spokes continuously improve product quality based on usage and consumer feedback.
\end{enumerate}

To evaluate whether the architecture truly breaks the flexibility-versus-control trade-off, we define three measurable outcomes. Let $U$ be the number of active monthly consumers for a data product portfolio, $F$ the median time required for a user to discover a fit-for-purpose asset, and $I$ the elapsed time from business question to validated insight. Improvement should be visible as an increase in $U$ and reductions in both $F$ and $I$. We define a normalized platform value score as
\begin{equation}
\label{eq:value}
V = w_u \frac{U}{U_0} + w_f \left(1 - \frac{F}{F_0}\right) + w_i \left(1 - \frac{I}{I_0}\right),
\end{equation}
where $U_0$, $F_0$, and $I_0$ are pre-transformation baselines and $w_u + w_f + w_i = 1$. This formulation captures the principle that platform success should be judged by tangible business consumption and discoverability, not only by internal platform activity.

In practice, $U$ can be derived from catalog access logs or query-engine audit trails, counting distinct teams or service principals that consume a product within the rolling window. $F$ is captured through clickstream telemetry on the catalog or conversational interface, measuring elapsed time from first search event to asset selection. $I$ is the most challenging to automate: practical proxies include ticket-based measurement from request to deliverable sign-off, or pipeline telemetry from first query to first downstream materialization. Baselines $U_0$, $F_0$, and $I_0$ should be established over a representative period before the architecture transition begins.

\section{Comparative Analysis}
\label{sec:results}

\begin{figure}[t]
  \centering
  \definecolor{f4gray}{RGB}{161,176,185}    
  \definecolor{f4gold}{RGB}{201,168,75}     
  \definecolor{f4teal}{RGB}{0,99,79}        
  \definecolor{f4navy}{RGB}{0,42,87}        
  \definecolor{f4grid}{RGB}{200,200,200}    
  \begin{tikzpicture}
  \begin{axis}[
    width=0.92\linewidth,
    height=8.5cm,
    ybar,
    bar width=18pt,
    ymin=0, ymax=3.75,
    ytick={1,2,3},
    yticklabels={Low, Medium, High},
    ytick style={draw=none},
    yticklabel style={font=\small, text=f4navy},
    ylabel={Qualitative assessment},
    ylabel style={font=\small\bfseries, text=f4navy},
    xtick={1,2,3,4},
    xticklabels={
      {\shortstack{Monthly product\\adoption}},
      {\shortstack{Time-to-find\\(inverse)}},
      {\shortstack{Time-to-insight\\(inverse)}},
      {\shortstack{Governance\\trust}}
    },
    xticklabel style={align=center, font=\small, text=f4navy},
    xmin=0.35, xmax=4.65,
    title={\textbf{Illustrative qualitative outcomes by organizational model}},
    title style={font=\normalsize, at={(0.5,1.05)}, anchor=south, text=f4navy},
    legend style={
      at={(0.5,0.98)}, anchor=north,
      font=\small, draw=gray!50,
      fill=white, fill opacity=0.95, text opacity=1,
      legend columns=3,
      column sep=10pt,
    },
    legend cell align=left,
    legend image code/.code={
      \draw[#1, draw=none] (0cm,-0.12cm) rectangle (0.35cm,0.12cm);
    },
    ymajorgrids=true,
    grid style={dashed, f4grid, line width=0.5pt},
    axis line style={f4navy},
    tick style={f4navy},
    clip=false,
  ]
    \addplot[fill=f4gray, fill opacity=0.9, draw=white, line width=0.6pt]
      coordinates {(1,2) (2,1) (3,1) (4,3)};
    \addlegendentry{Centralized platform}
    \addplot[fill=f4gold, fill opacity=0.9, draw=white, line width=0.6pt]
      coordinates {(1,2) (2,2) (3,2) (4,1)};
    \addlegendentry{Pure data mesh}
    \addplot[fill=f4teal, fill opacity=0.9, draw=white, line width=0.6pt]
      coordinates {(1,3) (2,3) (3,3) (4,3)};
    \addlegendentry{AI hub-and-spoke}
  \end{axis}
  \end{tikzpicture}
  \caption{Illustrative qualitative comparison of operational outcomes across three models, using low, medium, and high bands rather than measured scores. The proposed architecture aims to improve adoption and trust while reducing discovery and delivery latency.}
  \label{fig:outcomes}
\end{figure}
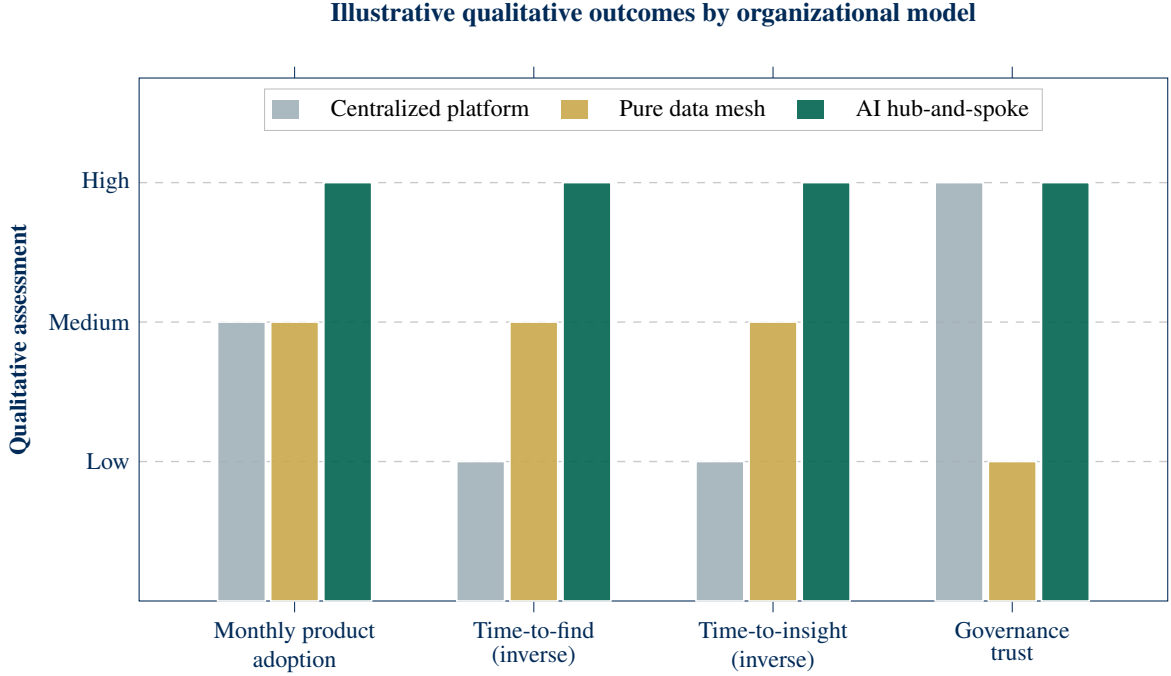

Because this paper is a design and synthesis study rather than a controlled experiment, our analysis is expressed as an operational comparison between three organizational regimes: centralized data platforms, pure data mesh, and the proposed AI-augmented hub-and-spoke model. Figure~\ref{fig:outcomes} summarizes the expected directional effects based on recurrent industry pain points and the capabilities described above.

The comparison yields four practical findings.

First, centralized teams often preserve control at the expense of throughput. They can apply standards consistently, but they struggle to model the semantics and urgency of every domain simultaneously. This leads to queueing behavior: the platform becomes orderly but slow. Pure mesh approaches reverse the trade-off, enabling local flexibility and faster iteration while allowing standard fragmentation, duplicated pipelines, and inconsistent metadata.

Second, the hub-and-spoke model improves on pure decentralization by retaining a durable center of gravity for standards and automation. The CoE remains responsible for cross-domain interoperability, shared semantics, policy publication, and platform reliability. Crucially, AI reduces the marginal coordination cost of these responsibilities. Metadata authoring, schema mapping, and quality rule generation are tasks that are too frequent to centralize manually and too important to leave entirely optional. Automation makes them scalable.

Third, conversational access changes who can benefit from the platform. Traditional catalogs assume technical literacy and familiarity with table structures, lineage graphs, and query languages. A chat interface linked to governed metadata lowers the activation energy for operational and business users. This matters because many enterprises possess substantial stores of underused data that are technically accessible but practically undiscoverable. The proposed architecture turns discoverability into a first-class product feature.

Fourth, the success criteria become clearer. Platform teams frequently report activity metrics such as jobs run, tables ingested, or tickets resolved. These are useful internal indicators but weak proxies for business value. The proposed evaluation framework instead prioritizes usage, time-to-find, and time-to-insight, which are closer to the underlying purpose of a shared data platform. A platform that is well governed but rarely used is not successful; likewise, a fast-moving domain platform that cannot be trusted for cross-enterprise decisions has failed in a different way.

To illustrate how the platform value score $V$ can operationalize these criteria, consider a hypothetical enterprise with baselines $U_0 = 200$ monthly active consumers, $F_0 = 45$~minutes median discovery time, and $I_0 = 5$~days median time-to-insight. Table~\ref{tab:value-score} shows plausible directional estimates under each regime with equal weights ($w_u = w_f = w_i = 1/3$). While the absolute numbers are illustrative rather than measured, the exercise demonstrates that $V$ captures regime differences in a single comparable metric and highlights the dimensions on which each model is strongest or weakest.

\begin{table}[H]
  \centering
  \caption{Illustrative platform value score under three organizational regimes.}
  \label{tab:value-score}
  \begin{tabular}{lrrrl}
    \toprule
    Regime & $U$ & $F$ (min) & $I$ (days) & $V$ \\
    \midrule
    Centralized platform & 220 & 40 & 4.5 & $\tfrac{1}{3}(1.10)+\tfrac{1}{3}(0.11)+\tfrac{1}{3}(0.10)=0.44$ \\
    Pure data mesh       & 260 & 35 & 3.0 & $\tfrac{1}{3}(1.30)+\tfrac{1}{3}(0.22)+\tfrac{1}{3}(0.40)=0.64$ \\
    AI hub-and-spoke     & 350 & 15 & 1.5 & $\tfrac{1}{3}(1.75)+\tfrac{1}{3}(0.67)+\tfrac{1}{3}(0.70)=1.04$ \\
    \bottomrule
  \end{tabular}
\end{table}

Table~\ref{tab:operating-model} maps responsibilities across the hub and the spokes. The model does not eliminate central work. It makes central work more focused on leverage: codifying policy once, publishing reusable abstractions, and instrumenting cross-domain quality so that autonomy can increase without loss of coherence.

\begin{table}[H]
  \centering
  \caption{Illustrative responsibility allocation in the proposed operating model.}
  \label{tab:operating-model}
  \begin{tabular}{|p{0.22\linewidth}|p{0.34\linewidth}|p{0.34\linewidth}|}
    \hline
    \textbf{Capability} & \textbf{Hub / CoE} & \textbf{Domain spokes} \\
    \hline
    \textbf{Platform engineering} & Lakehouse platform, catalog, policy engine, observability & Feedback on service gaps, domain-specific extensions \\
    \hline
    \textbf{Data product design} & Templates, minimum standards, contract schemas & Product backlog, semantics, service-level expectations \\
    \hline
    \textbf{Quality management} & Rule generation services, regression review, exception workflows & Approval of domain thresholds, remediation of failed checks \\
    \hline
    \textbf{Governance} & Global controls, access policies, stewardship council & Local stewardship, classification, evidence for exceptions \\
    \hline
    \textbf{Consumption enablement} & Search, chat interfaces, shared semantic layer & Business onboarding, domain-specific analytic narratives \\
    \hline
  \end{tabular}
\end{table}

\section{Discussion}
\label{sec:discussion}

The proposed model is intentionally skeptical of doctrinal purity. In practice, enterprises rarely possess the organizational maturity, engineering consistency, and product management discipline required for pure data mesh from the outset. Attempting to decentralize too aggressively can convert a queueing problem into a coordination problem. The surface appearance of autonomy improves, but the system remains fragile because teams reinterpret standards, postpone documentation, and produce uneven interfaces for consumers.

The hub-and-spoke architecture addresses this by distinguishing between \emph{where decisions are made} and \emph{where enabling capabilities live}. Domain teams should own the meaning, timeliness, and business fitness of their data products. A central hub should own the mechanisms that make those products legible, trustworthy, and discoverable at enterprise scale. This division is consistent with broader platform engineering guidance: reduce cognitive load on product teams by making the right path the easy path~\cite{skelton2019}.

AI is useful in this setting precisely because it can absorb documentation and review work that humans often defer. Still, its role must remain bounded. AI-generated contracts, metadata, and policy recommendations should be treated as accelerants subject to human approval, not autonomous governance actors. Poorly supervised automation can amplify existing quality issues or encode flawed semantic assumptions. The right design posture is therefore augmented governance rather than replaced governance.

A complementary dynamic reinforces the case for progressive spoke ownership. LLMs lower the barrier for domain practitioners to develop genuine cross-functional expertise, combining business knowledge with data engineering or data science capability that would previously have required years of specialised training. A commercial analyst who can author and validate a data pipeline, or a data engineer who can meaningfully engage with business definitions and quality expectations, reduces the volume of decisions that must be escalated to the CoE. Spoke teams with this multidisciplinary depth are better equipped to own their data products end-to-end---from ingestion through publication to consumer support---and the hub can correspondingly concentrate its effort on higher-leverage platform work rather than routine enablement.

The CoE can further operationalize this by publishing a curated library of codified skill definitions for AI coding assistants---machine-readable documents that a coding agent can retrieve and act on directly within the domain's development environment. Each definition covers a common enterprise source system and encodes the platform's preferred ingestion approach, approved languages and frameworks, compute utilization guidance, and reusable boilerplate scripts that teams adapt rather than build from scratch. A domain team whose coding assistant can locate and apply one of these definitions gains access to embedded platform expertise without depending on CoE bandwidth, allowing spoke teams to progress further along the maturity curve than their headcount alone would otherwise permit. This is not a replacement for platform investment, but it does mean that the ceiling on spoke maturity is rising faster than in previous eras.

Several limitations remain. First, the paper is prescriptive rather than empirical, drawing on published principles and recurring industry patterns rather than on a longitudinal deployment study. The illustrative values used for the platform value score $V$ are directional estimates, not measurements; validating them requires instrumented deployments across diverse organizational contexts. Second, regulated environments such as financial services and healthcare may require stricter approval loops, audit trails, and explainability guarantees than the model currently assumes. In such settings, AI-generated governance artifacts may need additional human-in-the-loop checkpoints and regulatory sign-off before promotion. Third, the effectiveness of conversational access depends on the maturity of metadata and authorization systems; natural-language interfaces cannot compensate for absent stewardship. If the underlying catalog is sparse or inconsistent, chat-based discovery will surface unreliable results. Fourth, the organizational transition from centralized or ad-hoc models to the proposed hub-and-spoke structure involves significant change management costs---including role redefinitions, incentive realignment, and training investments---that this paper does not quantify. Fifth, the AI components themselves introduce operational concerns: the central hub must manage LLM inference costs, and the security implications of AI systems that read sensitive metadata and schema information across domains. Sixth, the framework does not address the cold-start problem: organizations with minimal existing metadata or catalog infrastructure face a bootstrapping challenge before AI-enabled automation can deliver meaningful value.

Future work should validate the framework in live enterprise settings, ideally through longitudinal case studies that measure $V$ before and after adoption. Comparative studies across organizational sizes, data complexity levels, and regulatory environments would help characterize the boundary conditions under which the model is most effective.

\section{Conclusion}
\label{sec:conclusion}

Pure data mesh has proven attractive as a diagnosis of centralized bottlenecks, but much less reliable as a direct implementation strategy. The gap between theory and practice emerges because decentralization without leverage magnifies variation faster than it creates capability. A more durable path is to combine the principles of domain-oriented ownership with a modern lakehouse and an AI-augmented central hub that automates governance, standardization, and discoverability. The resulting hub-and-spoke model reframes the problem from choosing between flexibility and control to engineering the mechanisms that support both. Domain teams gain autonomy where they create the most value: semantics, prioritization, and local iteration. The hub preserves coherence where fragmentation is most costly: policy, interoperability, and platform reliability. If measured through product usage, time-to-find, and time-to-insight, this architecture provides a pragmatic route for scaling analytics and AI while keeping enterprise data FAIR and operationally trustworthy.

\bibliographystyle{unsrt}
\bibliography{references}

\section*{Declaration of Generative AI Use}

The authors used Claude Sonnet 4.6 (Anthropic) in the preparation of this work.
For code development, the tool was used to accelerate syntax generation, debugging, and boilerplate code structuring; the foundational logic and algorithmic designs were conceptualized by the authors.
For writing, Claude Sonnet 4.6 was used exclusively to improve language, grammar, and readability.
After using this tool, the authors thoroughly reviewed, edited, and verified all resulting text and code, and assume full responsibility for the accuracy and integrity of the final contents of this paper.

\end{document}